\documentclass{article}
\usepackage[paperwidth=8.5in, paperheight=11in, top=1.5in, bottom=1.5in, left=1in, right=1in]{geometry}
\usepackage{url}
\usepackage{multirow}

\usepackage{graphicx}
\usepackage{wrapfig}
\usepackage{authblk}
\usepackage[usenames,dvipsnames]{xcolor}

\newcommand{\y}[1]{#1}
\newcommand{\ny}[1]{#1} 
\newcommand{\ah}[1]{#1} 

\newcommand{\sctx}{secure-mode}
\newcommand{\csctx}{Secure-Mode}
\newcommand{\sm}{$\mu$TCB}
\newcommand{\smc}{The $\mu$TCB}
\newcommand{\smcc}{The $\mu$TCB}
\newcommand{\sak}{SAK} 

\date{}

\newenvironment{keywords}{
       \list{}{\advance\topsep by0.35cm\relax\small
       \leftmargin=1cm
       \labelwidth=0.35cm
       \listparindent=0.35cm
       \itemindent\listparindent
       \rightmargin\leftmargin}\item[\hskip\labelsep
                                     \bfseries Keywords:]}
     {\endlist}

\begin{document}

\title{Securing Smartphones: A Micro-TCB Approach}

\author[1]{Yossi Gilad}
\author[1]{Amir Herzberg}
\author[2]{Ari Trachtenberg}

\affil[1]{Computer Science Department, Bar-Ilan University}
\affil[2]{Electrical and Computer Engineering Department, Boston University}



\maketitle

\begin{abstract}
As mobile phones have evolved into `smartphones', with complex operating systems running third-party software, they have become increasingly vulnerable to malicious applications (malware). 
We introduce a new design for mitigating malware attacks against smartphone users, based on
\ah{
a small trusted computing base module, denoted \sm. 
The \sm\ manages sensitive data and sensors, and provides core
services to applications, independently of the operating system.}
The user invokes \sm\ using a simple {\em secure attention key}, which is pressed in
order to validate physical possession of the
device and authorize a sensitive action; this protects private information even if the device
is infected with malware. We present a \ny{proof-of-concept} implementation of \sm\ based on ARM's TrustZone, a secure execution environment increasingly found in smartphones, \ny{and evaluate our implementation using simulations}.
\end{abstract}


\section{Introduction}\label{intro}


The increasing popularity of cellular phones has steadily shifted personalized computing from wired desktop machines to wireless mobile devices. Accompanying this shift was a natural move to utilize mobile phones as `trusted digital agents'~\cite{journals/cacm/Herzberg03},
allowing users to receive private messages and perform mobile payments or other confidential tasks. Early mobile phones were simple, limited, and had proprietary implementations, leading to the impression that they were more secure than their multi-purpose desktop counterparts, on \y{which} users could install unchecked, and possibly malicious, third-party software.

\ah{
Today's cellular phones, however, have evolved into \emph{smartphones}: general purpose, complex, computer/sensor devices that employ multiple wireless communication technologies and are ever connected to both the Internet and their users. In effect, today's smartphones are a natural extension of a person's daily
life: texting/emailing friends, catching up on news/stocks, and managing bank accounts. Not only do they store passwords, account numbers, pictures, and videos that were freely saved on yesterday's phones, but they also place numerous sensors well within the comfort zone of their users.

It is therefore not surprising that the} evolution to smartphone devices arrived with a concomitant rise in exploits and attacks. The complex operating systems \ny{and variety of applications} in today's smartphones pose a rich attack surface that was not present in the earlier phones. Malicious applications steal user data, gain administrative privileges by exploiting software vulnerabilities~\cite{3gppstack} and steal user credentials by faking legitimate notifications~\cite{notificationphishing}. Worse yet, the vulnerability/patch cycle for mobile devices is often long, due, in part, to the potential need to modify the code of several phone manufacturers~\cite{conf/uss/VidasVC11a}.  Finally, \ah{many smartphone users are non-technical, making security an even greater challenge.}

A successful security architecture must rely as little as possible on the user's skill-set. It should be simple \ah{and small} enough that its execution paths can be evaluated for security, and yet sufficiently elemental to support a variety of services. In this paper we propose one \ah{such} security architecture, which uses the phone's physical proximity to its user as a basis for authentication.

\ny{
\paragraph{Threat Model.}
This work considers attacks by malware applications.  The standard defense against such attacks is {\em application sandboxing}, wherein a permission system offers applications a variety of approved privileges to resources, ranging from sensor access to the ability to display notifications. These privilege-restrictions limit the potential damage of malicious or exploitable applications.

However, the privilege systems \ah{fail if} an application `breaks out of jail' by exploiting vulnerabilities in the (complex) smartphone operating system or a privileged third-party application. Indeed, even security-focused application-level defenses (e.g., ~\cite{safeslinger}) may fail when the operating system itself is compromised by malware. Under such conditions, privileged malware can read files, monitor the keyboard, record the microphone, take pictures and learn the user's location, all posing a significant threat to users' privacy.

In order to function even under operating system compromise, our security design} \ah{is based on 
a \emph{minimal} Trusted Computing Base (\sm) that supports a few necessary operations. 
We assume that the \sm\ implementation executes in a dedicated environment and is small enough to be thoroughly validated, we thus preclude attacks on this component. Our security design further assumes that the user currently possesses the phone, and thus we preclude physical attacks. 
}

\subsection{Design Overview}
We integrate a simple and minimal trusted component into the smartphone, and migrate key security-oriented services from the base operating system~(e.g., Android\texttrademark) to this component. In a way, this component resembles the micro-kernel approach to \ah{secure} operating system (OS) design, in which the OS is reduced to a minimal kernel and most functionalities are implemented outside the kernel; we thus name our trusted component a
\emph{Micro Trusted Computing Base~(\sm)}.


The core of our design is the separation between the smartphone's complex OS and the
\sm, which allows applications that execute over the smartphone's OS to access its services using a programming interface. 
\ah{In addition,} 
when the user presses a dedicated button, called the {\em Secure Attention Key (\sak)}, 
the device enters a special \ah{{\em \sctx}.  In \sctx,} sensor signals and hardware interrupts are first processed within the \sm\ environment, and only then possibly relayed to the OS. This allows \sm\ to take control of peripherals, such as the screen, and interact with the user securely even if the OS or other phone subsystems are compromised.  
By pressing the \sak, the user validates that she physically possesses the device, thereby
distinguishing between automated or remote instructions and user commands.

%
%

\y{
We demonstrate the applicability of our design with two fundamental examples, secure messaging and mobile-payment applications\ah{; both} execute within the smartphone's OS but base their security on \sm\ services. As a proof-of-concept, we implement our \sm\ design within the ARM TrustZone security environment \ny{and provide a performance evaluation using ARM's device simulator}.
} 
\section{Related Works}\label{rworks}

\subsection{Trusted Physical Interfaces}
\ny{A secure path between the user and a trusted system component is a key requirement of a usable security architecture, and this is particularly true for smartphones~\cite{conf/trust/VasudevanOZNM12}, which are widely used \ah{by naive} users. The concept of using a dedicated secure attention key (SAK) to establish such a path was introduced in~\cite{SAKFirst} for opening a secure login-shell. This design is incorporated in modern operating systems, for example, Windows' Ctrl+Alt+Del combination, but is typically limited to login or task management, and can be affected by a compromised operating system.

Our design adapts this idea of a dedicated button to \ah{smartphones.  We secure private user data, such as login passwords, files, and messages, and enable secure management of sensitive sensors.}
Finally, we provide a programming interface that allows complex (external) applications to delegate their security building-blocks to the simple \sm\ core. 
%

%

Among a variety of related works, McCune et al. considered the problem of user-observable path verification~\cite{turtles}, where the user receives a clear and unspoofable feedback that she communicates securely with a trusted endpoint. To this aim,~\cite{turtles} suggests using a LED indicator connected to the path endpoint. Our design includes a similar indicator, attached to the \sm, which signals when the device is in \sctx\ and the user can interact securely with the \sm.
However, we do not solely rely on passive indicators, combining additional mechanisms to create a `defense in depth'-like architecture: we train the user to press the \sak\ prior to every sensitive action (ensuring that the device is in \sctx), as well as force the user to enter \sctx\ to perform certain sensitive operations (such as signing documents).
}

\subsection{Trusted Computing Base (TCB)}
TrustVisor~\cite{conf/sp/McCuneLQZDGP10} is a small TCB, designed for personal-computers, that allows applications to execute security-sensitive codeblocks (e.g., signature verification).  Complementarily, the Trusted Language Runtime (TLR)~\cite{Santos:2011:TLR:2184489.2184495} architecture provides similar capabilities for smartphones. While TrustVisor and TLR provide important security guarantees, they are strictly designed to secure \emph{computations} and not interactions with the user; for example, a malware that replaces the keyboard device driver could, under both systems, record or modify the user's keystrokes.


TrustVisor has been extended to create a trusted path between an input/output (I/O) device and an application~\cite{conf/sp/ZhouGNM12}. The design of~\cite{conf/sp/ZhouGNM12} requires that each application implement its own trusted I/O device functionality (as part of the TCB); in contrast, our \sm\ provides trusted services to untrusted applications (through a restricted programming interface). This makes it easier to code applications, but also makes our design less flexible (e.g., only the manufacturer can introduce new TCB primitives).  We believe that codability is essential for today's constantly growing smartphone application markets: only the most sophisticated developers can be expected to properly write and {\em thoroughly verify} their TCB application code.  In addition, the verification of the trusted path in~\cite{conf/sp/ZhouGNM12} makes use of an additional hand-held device, which can be an impediment to consumer adoption; our design {\em embeds} a trusted path between the user and \sm, which the user can enable by pressing \ah{the \sak}. 
This allows us to 
incorporate the verification indicator into the smartphone, by placing it under the control of \ah{the \sm.}



The trusted sensors architecture of~\cite{conf/mobisys/LiuSWR12} uses a TCB to provide reliable sensor readings for applications, 
and Raj et al.~\cite{conf/wmcsa/RajSWP13} show how a TCB could monitor network traffic and attest for its volume. These architectures provide important services, but do not intrinsically prevent malware applications from obtaining sensor readings or accessing sensitive files; they are thus complementary to our design.

\section{The Micro-TCB Design} \label{design}
In this section we present the \sm\ architecture and its interaction methods with the user.


\subsection{Architecture}
The \sm\ platform has two important properties. First, it provides a code-execution environment that is separated from the smartphone's OS, allowing \sm\ to remain secure even when the OS is compromised. Second, \sm\ code running over this platform \y{can handle or filter} signals from sensors and peripherals \emph{before} they are handled by the OS. This property allows privileged access to the secure attention key and touch-screen signals, and enables \sm\ services to confidentially interact with the user (see next subsection). 

\y{\smc\ is minimal by design \y{(in order to avoid vulnerabilities)}, and it thus relegates all but a few security-oriented services to the smartphone OS; these services, described in Section~\ref{services}, are accessible to applications running on smartphone's OS via a restricted API. Though minimal, \sm\ services allow implementation of complex and fundamental applications, as we show in Section~\ref{applications:s}.} 

\begin{figure}[t]
	\centering
	\includegraphics[width=0.85\textwidth]{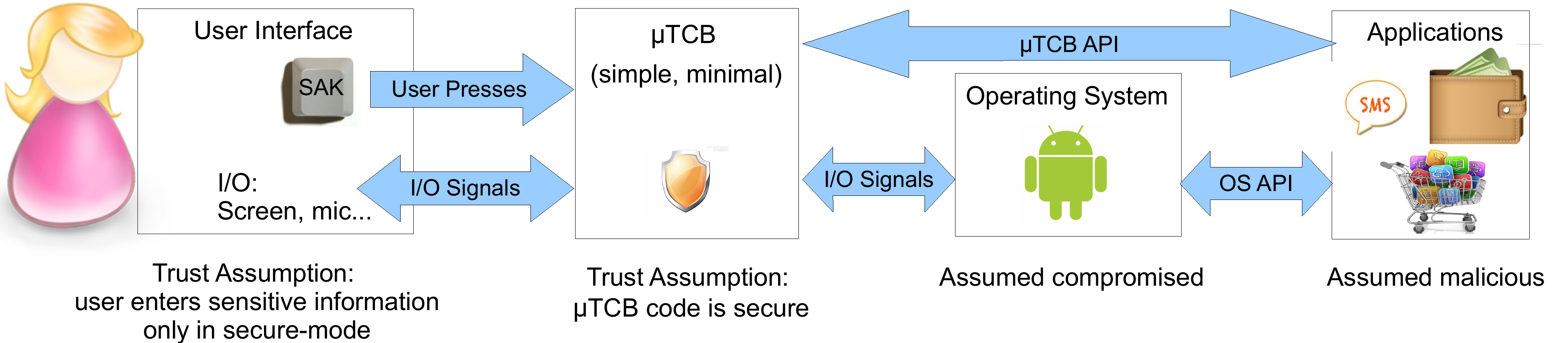}
	\caption{\smcc\ Architecture, Components and Trust Assumptions.\protect\footnotemark}
	\label{fig:highlevel}
\end{figure}

Figure~\ref{fig:highlevel} illustrates our architecture at a high level \y{and the system's `trust assumptions' in participating entities}, implementation details will be covered in Section~\ref{impl}.

\subsection{The Secure Mode and Attention Key} \label{design:sak}
\smc\ has a dedicated Secure Attention Key (\sak) that is not visible to the smartphone's OS. When the user presses the \sak, she triggers an interrupt that invokes an interaction with the \sm\ system; we call this interactive mode {\em \sctx}. The \sm\ presents \y{to the user} a menu of simple services \y{as well as pending requests from applications, we describe those services and requests in Section~\ref{services}.
}

\footnotetext{The Android robot is reproduced from work created and shared by Google and used according to terms described in the Creative Commons 3.0 Attribution License.}

\y{Pressing the \sak\ {\em ensures} that the smartphone is in \sctx, namely, the \sak\ is a `one-way button'. 
In order to exit the \sctx, the user presses a virtual `Exit' button in the \sctx\ menu. The smartphone also automatically exits \sctx\ after idle timeout (e.g., five minutes), this is to avoid confusing the user regarding the smartphone's mode after a long idle period.
}

When in \sctx, \y{\sm\ temporarily handles signals of the touch-screen and filters those of other sensors, blocking them from reaching the smartphone's OS} so as to ensure that no application, not even one that escapes `out-of-sandbox', may view or manipulate the screen to leak private user inputs.



\paragraph{\csctx\ Indicator.}
A key requirement of our system is that users receive a clear and unspoofable indication when their smartphone is in \sctx. 
We identify two cases where such indication is crucial: (1) the user enters secret data, (2) the user disables a sensor. In the first case, an application that tricks the user will learn the secret data; in the second case, a sensor that remains active might expose sensitive information.
To this aim, a LED indicator is connected to the \sm\ and lights only when the device is in \sctx\ (the LED is inaccessible to the smartphone's OS).

\paragraph{Forcing and Training Ceremonies.}
Usability studies suggest that users often do not notice passive security indicators; to improve security, we use {\em forcing and training ceremonies~\cite{conf/ndss/KarlofTW09,conf/esorics/HerzbergM11}}. 

First, the user's private data and keys are only accessible to \sm, {\em forcing} the user~\cite{conf/ndss/KarlofTW09} to press the \sak\ (and enter the `real' \sctx) in order to use them, e.g., to sign documents with her private key.

Second, we support a {\em training mechanism}, shown to improve indicator awareness~\cite{conf/esorics/HerzbergM11}. 
We assume an initial period where the smartphone is infection-free, and simulate a spoofing attack by occasionally neglecting to light the indicator LED when the smartphone enters \sctx. Any action by the user without re-pressing the \sak\ (i.e., to ensure that the device is in \sctx\ and indicator turns on) results in a negative feedback, e.g., the LED blinks and \sm\ notifies the user to notice the~indicator.

\section{Services and Interfaces} \label{services} 
\smc\ provides security-oriented services to the user as well as to applications running on the smartphone's operating system. These services are available through restricted user and application interfaces. 

\begin{figure}[t]
	\centering
	\includegraphics[width=.95\textwidth]{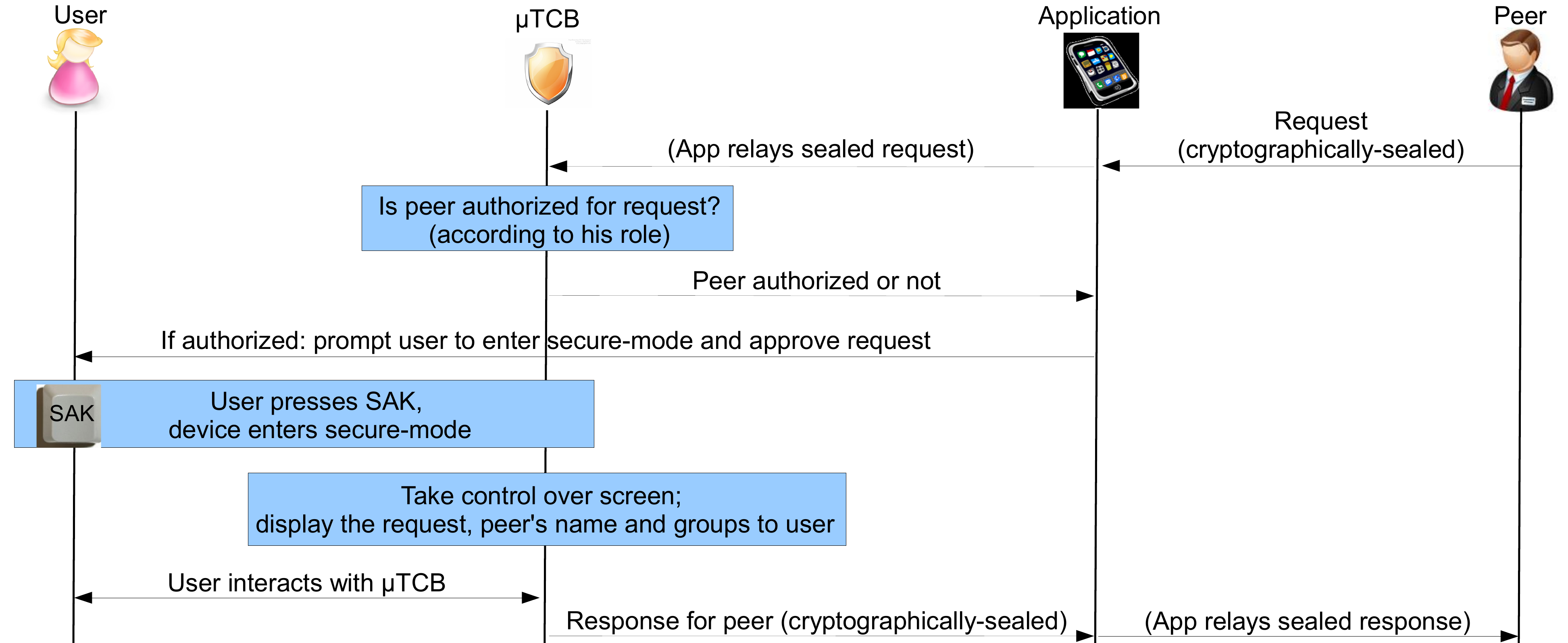}
	\caption{Role-Based Access Control API Scheme.}
	\label{fig:apischeme}
\end{figure}

Our application programming interface (API) embeds \emph{role-based} access restrictions, as illustrated in Figure~\ref{fig:apischeme}, to allow applications to access \sm\ services only \emph{on behalf of} a remote peer \y{(representing a remote entity like a cloud service, contact person or online store)} that has permission to use the service. In such cases, \sm\ uses a \y{public} key assigned to the peer to cryptographically seal/unseal data for/from the calling application; in essence, role-based access control allows the application to only relay, but not read or modify, the sensitive information \ny{and, therefore, prevents exposure of such information in the case that the application is malicious}. 


We next present the services and interfaces that \sm\ offers, organized into four categories: authorization, confidentiality, non-repudiation and sensor control. 

\subsection{Authorization of Peers, Roles and Groups} \label{serv:auth}
\smc\ restricts sensitive operations, such as those that disclose private information, to authorized peers. Each peer is represented by a name assigned to a public key. The peer is authorized for permissions in two logically orthogonal classes: {\em groups} and {\em roles}.

Groups restrict the distribution of information, and can be modified or created (by a certificate authority or the~user); for example, work-related documents may be restricted so that they are only sent (under encryption) to the group of employees of the company.

Roles, on the other hand, restrict the actions permitted by peers.  There are exactly three~roles:
\begin{description}
\item [Contact:] this role enables a peer to send/receive encrypted and authenticated information to/from the~user. 

\item[Signatory:] peers with this role can send signed documents and request a counter-signature from the user. 


\item [Certificate authority (CA):] peers with this role are trusted to identify and authorize other peers. A CA is restricted to authorizing peers of specific groups (e.g., a company's CA can only authorize its employees). Credentials of well-known CAs are installed to \sm\ before user-adoption, e.g., by the manufacturer.
\end{description}


\paragraph{User Interface and API.}\label{authorization:ui} The user may authorize new peers. However, this management option is risky and should typically be used only by system administrators to authorize a local CA; as such, this option may be `locked' with an administrative password.

%
%
The authorization service is inaccessible to applications, to avoid exposing the user to social engineering attacks (e.g., where the malware authorizes a peer with similar name to the user's bank).


\vspace{-2pt}
\subsection{Confidentiality}\label{serv:repository}
\vspace{-2pt}
Smartphones often store sensitive data, ranging from login credentials, to private pictures and messages; \sm\ provides means of ensuring the confidentiality, authenticity and integrity of such data, and an infrastructure for distributing it among {\em contact} peers. 

More precisely, \sm\ includes a {\em private repository}, which is essentially a secure file system \y{(inaccessible to the~OS)}. Some of the user's files 
in the repository, and in particular her {\em private key}, are encrypted with a user-provided credential such as a password; \ny{the user's credential is never saved, so that she must be directly involved to access this information, helping to mitigate the risk of identity theft from a stolen phone.}

\smc\ can also encrypt and authenticate data for transmission to contact peers. In this case, the sender's \sm\ uses the recipient's public key to encrypt two symmetric keys: one to encrypt and the other to authenticate the information. The encrypted keys are signed using the sender's private key, which allows the recipient to validate the authenticity and integrity of messages from the sender. \y{We encapsulate messages using the ``encrypt-then-MAC'' paradigm, which permits verification of the message's authenticity prior to decryption, to quickly discard spoofed messages.} 



\paragraph{User Interface and API.}
The user can access the private repository as a general purpose file system. Each file is associated with an access list of contact peers, identified by their names or groups, who may receive it.


Applications can use the \sm\ to handle private information on~behalf of authorized {\em contact} peers (following Figure~\ref{fig:apischeme}). The API functions that access the confidentiality service state the intended recipient or the sender of the data, and \sm\ validates that he is an authorized contact;
the validation result is returned to the application. If the peer is authorized, then the application prompts the user to press the \sak\ and enter \sctx\ to interact with \sm. The user can view the peer's name and groups while in \sctx. The two confidentiality-service API functions operate as~follows:

{\sf Request-Data(\textit{recipient})}. \smc\ requests the user to enter a message, select a file from the private repository (that {\sf recipient} is authorized to read) or send sensor-readings (e.g., camera picture, location coordinates); \sm\ then encapsulates the information as described above and returns the cryptographically-sealed data to the application.

{\sf Display-Message(\textit{sender}, \textit{message})}. \smc\ displays {\sf message} along with {\sf sender}'s name and~groups.


\subsection{Non-Repudiation}\label{serv:repudiation}
\vspace{-1pt}
\smc\ allows the user to sign documents presented by {\em signatory} peers, possibly after the user completes her personal details (e.g., name, identification number, etc.), while ensuring that neither the user's private key nor her personal details leak to OS applications. 

\ny{Each document has a type (e.g., banking, commerce, etc.) and some important types are~reserved and may be presented only by peers in specific groups (e.g., only a bank may present a banking~form).}

\vspace{-2pt}
\paragraph{User Interface and API.}
\smc\ stores the documents that the user has signed, and allows the user to later view those documents. 
%
Applications can request the user to sign or display documents on behalf of \textit{signatory} peers via the following API functions (that implement the scheme in Figure~\ref{fig:apischeme}):

{\sf Request-Signature(\textit{recipient}, \textit{document})}. When an application calls this function, \sm\ validates that {\sf recipient} is an authorized signatory peer, that {\sf {document}} is signed by {\sf recipient}, \y{and that {\sf recipient} may request signature for this document type}. 

When the user presses the \sak\ and places the smartphone in \sctx, she can view the document and possibly complete required personal details; if she approves the signature request, then \sm\ uses her private key to counter-sign the document (the user signs the completed document and the original signature~of {\sf {recipient}}). The application receives the user's signature and personal details encrypted with {\sf recipient}'s public~key.

{\sf Display-Signed-Doc(\textit{sender}, \textit{document})}. This function is similar to the previous one, but only uses \sm\ to display {\sf sender}'s signed {\sf document} to the user (i.e., the user does not counter-sign~it).

\y{
\subsection{Sensor Control} \label{sensors}
\vspace{-1pt}
The sensor-rich smartphone allows privileged applications to record, take pictures and locate the user {\em at any time}. 
\ah{
Existing smartphone} operating systems typically permit users to turn off some sensors (such as the GPS) for power conservation, but other sensors (such as the microphone or camera) are not directly controllable. Moreover, on an infected device any of these sensors may be activated without indication to the user. 

In our architecture, the sensors signals are routed to the OS via \sm, allowing it to completely block their signals from the OS. 

\paragraph{User Interface and API.}\label{sensors:ui}
The user can block or activate the sensors.
%
Applications can request the user to {\em temporarily} enable a sensor via the {\sf Enable-Sensor(\textit{sensor})} function. When an application calls this function, \sm\ validates that {\sf sensor} is currently blocked (or will be re-blocked soon, see below).

When the user presses the \sak, she can approve activating the sensor for a short period of time (e.g., up to an hour), or choose to automatically discard activation requests for some time period. \smc\ notifies the application of the user's decision and its timeout, allowing the application to again request to activate the sensor when the decision expires.
}

\section{Applications} \label{applications:s}
\ny{In this section we describe example applications that illustrate how \sm\ and its API allow
implementation of key security-demanding applications.} 



\subsection{Messaging} \label{sms}

\begin{figure}[t]
	\centering
	\includegraphics[width=1\textwidth]{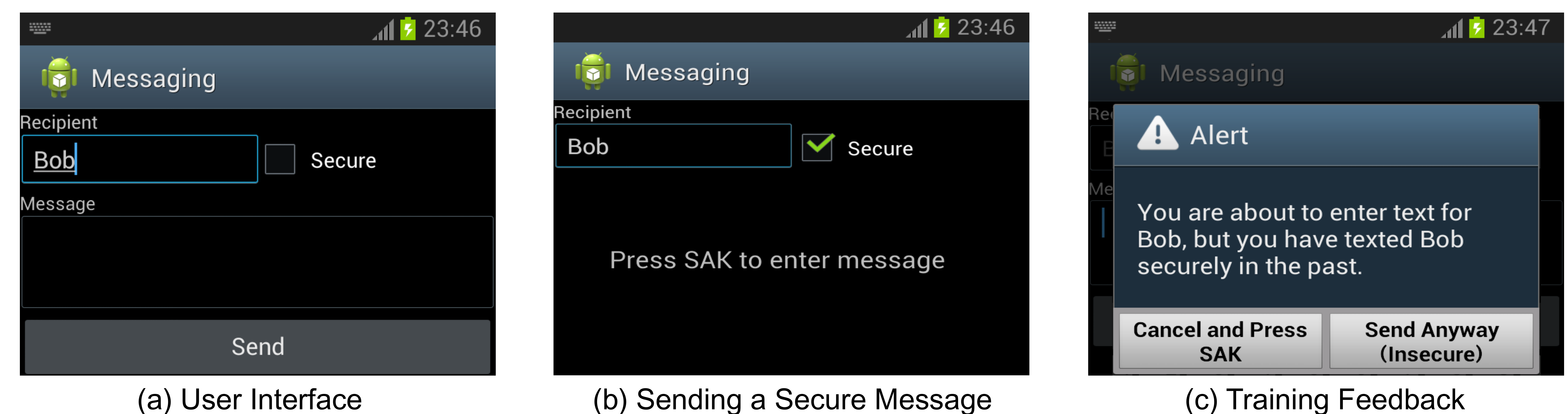}
  \caption{Secure Messaging. User interface and training feedback \ny{(in virtual machine)}.}
  \label{fig:screenshot}
\end{figure}

The secure-messaging application includes a check-box where the user can mark messages as `secure', see Figure~\ref{fig:screenshot}~(a). In order to send a secure message, the application retrieves the recipient's certificate (which includes his public key, groups and role) by requesting it from the recipient's device, e.g., via SMS or network interface \y{(the certificate is signed by a CA, and signature validation performed by \sm)}. 

When the user sends a secure message, the application calls the {\sf Request-Data} API function, and prompts the user to press the \sak\ and enter a message for that recipient, as in Figure~\ref{fig:screenshot}~(b). \smc\ returns to the application the {\em cryptographically-sealed} message, and the application then sends the message to the recipient. 

When the application receives an encapsulated message, it calls the {\sf Display-Message} API function and prompts the user to enter \sctx\ to read the message.

In order to encourage users to incorporate the secure functionality, the secure option is set as default for communication with contacts to whom the user had sent/received a secure message before. If the user changes this configuration and attempts to send a plaintext message to such contact, the application alerts the user, as in Figure~\ref{fig:screenshot}~(c).








\subsection{Commerce and Payments} \label{applications:payments}

We next present a mobile payment protocol that integrates the \sm\ to mitigate the risks of identity theft and exposure of payment credentials. Our description uses the example of a stock trading application for sake of exposition. We assume that prior to running the protocol, the user has registered her public key with her bank.



\paragraph{Protocol.}
In the {\em offer} phase, the user requests to buy a particular stock using a smartphone application. The application might find several available brokers, all of which should be \emph{signatory} peers approved by trusted CAs (e.g., banks), and it presents to the user the information required to decide on a particular offer (e.g., the broker's commission). The application calls the {\sf Request-Signature} API function, which provides to \sm\ the identity of the selected broker and his signed~offer.

In the {\em payment} phase, the user enters \sctx\ and views the broker's offer; if she decides to accept it, then she enters the secret credential (e.g., password) for decrypting her private key file in the repository (see Section~\ref{serv:repository}). \smc\ uses \ah{the} 
private key to sign an order after the user completes relevant personal details (e.g., account number). The signed order is cryptographically sealed using the broker's public key and the cipher-text is returned to the application which then relays it to the broker.


Finally, in the {\em confirmation} phase, the broker decrypts the user's response and clears the payment with her bank. The broker sends a signed receipt to the application, which then calls the {\sf Display-Signed-Doc} API function and notifies the user to enter \sctx\ to view the receipt. If the user does not receive a receipt in a timely fashion, she can notify her bank and revoke the transaction~(this may be automated).

\paragraph{Security Benefits.} This payment protocol improves typical payment applications since it mitigates the risk for identity theft: the user does not expose her payment credential (i.e., private key) to the bank, broker or the application (cf. to credit-card payments). \y{Moreover, even if her smartphone is lost or stolen, a user-chosen credential is required to decrypt her private key, further reducing the risk for impersonation or fraud.} 



\section{Implementation} \label{impl}
\ny{In this section we describe our Proof of Concept (PoC) implementation of \sm\ over ARM TrustZone, a trusted code execution platform already available in some of today's mobile devices.} TrustZone separates between the {\em normal-world~(NWorld)}, where the rich OS and applications execute, \ny{and} the {\em secure-world~(SWorld)}, where the~\sm\ resides.

\y{
We implemented a {\em proof of concept} (PoC) prototype to verify that implementation of our design over TrustZone is feasible. Our PoC 
implements the \sm\ services and provides an API (in form of a library) for applications running in NWorld; our TCB has less than 8000 lines of source code and is much smaller than the phone's OS kernel. We note that our PoC does not modify the smartphone's OS to provide applications access to TrustZone and the \sm\ APIs. A full implementation would need to make this modification for a commercial product, but for feasibility\ny{-testing} purposes it is sufficient to execute simple applications directly over NWorld.
}

\paragraph{System Boot.}
\smc\ (running in SWorld) initializes prior to the smartphone's OS (running in NWorld). The TrustZone platform allows \sm\ to specify which peripherals' interrupts should be handled by the smartphone's OS, and which by \sm. Initially, the only I/O devices handled by \sm\ are the \sak\ and the LED indicator, all other peripherals are handled in NWorld by the smartphone's OS. 

\paragraph{Switching to \csctx\ and Interrupt Masking.}
\ny{
When the user presses the \sak, \sm\ registers itself to handle interrupts from the touch-screen, which is required to interact with the user while the smartphone is in \sctx\ (this is changed back when the user exits \sctx). If the user requests to block a sensor for privacy reasons (e.g., microphone, camera), \sm\ masks its interrupts. The only full peripheral-device driver required to deploy in \sm\ is, therefore, that of the touch-screen.

Such a scheme efficiently handles interrupts in a TrustZone based implementation of \sm. However,
its implementation is difficult: it requires careful bookkeeping of the touch-screen's state which should be restored when the device exits \sctx, and handling edge cases where the OS is currently using the peripherals that \sm\ needs to block. We leave addressing these issues in our prototype for future work, and avoid them in our evaluation below by using simple applications that execute directly over NWorld.
}


\paragraph{Architecture Overhead Evaluation.}
\ny{In order to evaluate our implementation, we utilized the ARM FastModels product that simulates an ARM device with the Cortex-A9 processor that is common in smartphones. We run the simulator on a PC with an Intel Core 2 Duo processor and 4GB of RAM, resulting in a wait time of roughly 80 seconds to load the simulation. In order to evaluate the overhead of \sm, we measure the number of additional processor cycles while performing different operations (system boot, API calls and interactions with the user) and use these measurements to estimate execution time on the Cortex-A9 processor.}

\ny{The overhead introduced to the smartphone boot process is minimal: the \sm\ only initializes the interrupt handlers, and switches to NWorld to continue the normal boot sequence of the smartphone; the \sm\ initialization requires 1600 processor cycles (less than 3$\mu$s on the Cortex-A9 processor).  

The cost of switching between the normal and secure worlds is roughly 3000 cycles, or 5$\mu$s on the \mbox{Cortex-A9}. However, it is only `paid' when an application invokes the \sm\ via an API call or when the user presses the \sak\ to enter \sctx, and not during the typical operation of the smartphone: When not in \sctx, all I/O interrupts (except for the \sak) are handled in NWorld by the smartphone's OS, and therefore do not require (expensive) switching between worlds.

Handling the \sak\ has the highest overhead, since the \sm\ `takes over' the screen and displays the menu to the user, but requires less than \ny{$10^6$ cycles} or 2ms on the Cortex-A9, which is sufficiently short to remain unnoticeable to the user.
}

\section{Conclusions}\label{conclude}


Cellular phones have evolved to `smartphones', complex devices that are much more vulnerable to malware. This is especially alarming considering their widespread use and access to sensitive information, sensors data, video/audio feeds and the like.

In order to secure the complex smartphones, we have proposed to delegate critical security services from the smartphone's OS to a small and separated \sm\ module. \ny{Our design integrates the established concepts of minimal TCB, trusted physical interfaces, user training and forcing ceremonies within the context of smartphones.} We have demonstrated the use of \sm\ and \sak\ for two smartphone applications: messaging and mobile payments. Finally, we have presented an implementation of \sm\ by utilizing TrustZone, an existing mobile-phone trusted computing platform, and evaluated its performance using \ny{ARM's device simulator}.

\ny{
During this work, we found that although modern OSs are made aware of underlying hardware security platforms, it may be challenging to use some of the platforms' features in practice (e.g., TrustZone's interrupt-handlers remapping). We hope that our work will motivate OS vendors to improve their support for such platforms in the face of increasing threats to users through their smartphones.
}




\y{
\subsection{Future Work} \label{future}
Several significant steps are needed to bring our PoC implementation into a generally usable
system.  To deploy it with a fully-featured OS, the OS needs to be patched to allow efficient remapping of interrupt handlers and expose the \sm\ API library to applications.  The simulator code must also be ported to smartphone hardware, \ah{a nontrivial engineering challenge.}
Finally, a usability study with deployment on fully-functional smartphones is important in order to evaluate and improve the proposed security indicator, user interfaces and the interaction protocols between the user and \sm. 

This work also leaves several questions open for research. For one, it is unclear how to best present pending requests from applications to the user in \sctx.  It is also interesting to consider alternate \sctx\ indicators, and effective feedback methods to train users to notice such an indicator. Finally, it would be interesting to leverage the smartphone's sensors to derive keys without relying on text passwords.





}

\section*{Acknowledgments}
We thank Idan Warsawski for his contribution to an early Arduino-based implementation of this work.

This work was supported by grant 1012910 from the National Science Foundation (NSF), division of computer and network systems; grant 1354/11 from the Israeli Science Foundation (ISF); by the Check Point Institute for Information and Security (CPIIS); and by the Ministry of Science and Technology, Israel.
\bibliographystyle{unsrt}
\bibliography{cellsec}

\end{document}